\documentclass[
amsmath,
prd,nofootinbib,floatfix,12pt,
]{revtex4}

\newcommand\dbar{\overline d}
\newcommand\ubar{\overline u}
\newcommand\ebar{\overline e}
\newcommand\nubar{\overline \nu}
\newcommand\beq{\begin{eqnarray}}
\newcommand\eeq{\end{eqnarray}}

\allowdisplaybreaks
\usepackage{setspace}

\begin{document}
\renewcommand{\theequation}{\arabic{section}.\arabic{equation}}
\title{\Large%
Non-universal gaugino masses from non-singlet $F$-terms in 
non-minimal unified models} 
\author{Stephen P. Martin}
\affiliation{Department of Physics, Northern Illinois University, DeKalb IL 60115}
\affiliation{Fermi National Accelerator Laboratory, P.O. Box 500, Batavia IL 60510}


\begin{abstract} 
\baselineskip=18pt
In phenomenological studies of low-energy supersymmetry, running gaugino 
masses are often taken to be equal near the scale of apparent gauge 
coupling unification. However, many known mechanisms can avoid this 
universality, even in models with unified gauge interactions. One example 
is an $F$-term vacuum expectation value that is a singlet under the 
Standard Model gauge group but transforms non-trivially in the symmetric 
product of two adjoint representations of a group that contains the 
Standard Model gauge group. Here, I compute the ratios of gaugino masses 
that follow from $F$-terms in non-singlet representations of $SO(10)$ and 
$E_6$ and their sub-groups, extending well-known results for $SU(5)$. The 
$SO(10)$ results correct some long-standing errors in the literature.
\end{abstract}

\maketitle
\tableofcontents
\baselineskip=15.55pt

\newpage
\setcounter{footnote}{1}
\setcounter{page}{1}
\setcounter{figure}{0}
\setcounter{table}{0}

\section{Introduction\label{sec:intro}}
\setcounter{footnote}{1}
\setcounter{equation}{0}

The phenomenology of the Minimal Supersymmetric Standard Model (MSSM) 
\cite{MSSMreviews} has most often been discussed within the confines of 
the minimal supergravity framework in which gaugino masses and scalar 
squared masses are assumed to be unified as running parameters at a very 
high renormalization scale. This scale is usually taken to be near the 
apparent unification of the gauge couplings at about $2 \times 10^{16}$ 
GeV. However, it is well-known that qualitatively different types of 
models can arise if one abandons one or both of the gaugino mass or 
scalar mass unification assumptions. The dominant production and decay 
processes at colliders can be quite different and new mechanisms for 
obtaining the observed amount of cold dark matter become available in 
large regions of parameter space. Moreover, such models can be motivated 
by the fact that the supersymmetric ``little hierarchy" problem can be 
ameliorated if the gluino mass parameter is significantly smaller than 
the wino mass parameter at the unification scale 
\cite{KaneKing}-\cite{compressed}.

This paper is concerned with a mechanism for obtaining gaugino mass 
non-unification that can work even if the gauge couplings unify into a 
simple gauge group like $SU(5)$, $SO(10)$, or $E_6$. In general, gaugino 
masses in supergravity can arise from non-renormalizable dimension-5 
operators involving $F$-terms that get a vacuum expectation value (VEV):
\beq
{\cal L} &=& -\frac{F^{ab}}{2 M_{\rm Planck}} 
\lambda^a \lambda^b + {\rm c.c.}
\eeq
Here $\lambda^a$ is the two-component field for the gaugino, and the 
indices $a,b$ run over the adjoint representation of the gauge group. The 
resulting gaugino mass matrix is $\langle F^{ab} \rangle/M_{\rm Planck}$. 
The supersymmetry-breaking order parameter $\langle F^{ab} \rangle$ must 
transform as a singlet under the Standard Model gauge group $SU(3)_C 
\times SU(2)_L \times U(1)_Y$. In general, it can take independent values 
for the gluino, wino and bino, the gauginos corresponding to the three 
subgroups. However, one can make non-trivial predictions 
\cite{Ellis:1984bm}-\cite{Anderson:1996bg} in any model in which the 
Standard Model gauge group is embedded in a larger symmetry group,
like a supersymmetric Grand Unified Theory (GUT) \cite{SUSYGUTS}. A 
non-zero contribution to the gaugino masses can arise from $F^{ab}$ fields 
transforming in a representation in
the symmetric part of the direct product of the adjoint 
representation of the unified group with itself.

The simplest example of this mechanism is the case of a GUT model based on 
$SU(5)$ \cite{SU5GUT}. There, one has\footnote{All group theory facts and 
notations followed in this paper can be found in the review article 
\cite{Slansky}.}
\beq
({\bf 24} \times {\bf 24})_S = {\bf 1} + {\bf 24} + {\bf 75} + {\bf 200}.
\eeq
Each of the four irreducible representations on the right 
contains exactly one Standard Model singlet, so that each makes a unique 
prediction for the pattern of contributions to the bino, wino, and gluino 
mass parameters $M_1$, $M_2$, and $M_3$ at the breaking scale. The overall 
scale of each $F$-term VEV is an arbitrary input parameter, but the 
ratios of contributions to $ M_1$, $M_2$, and $M_3$ are rational numbers 
fixed by the group theory. They were obtained in 
refs.~\cite{Ellis:1985jn},\cite{Anderson:1996bg}, and are listed in
Table \ref{table:SU5}.
\begin{table}
\begin{tabular}[c]{|c|c|}
\hline
\phantom{X}$SU(5)$\phantom{X} & $M_1$ : $M_2$ : $M_3$ \\
\hline\hline
{\bf 1} & 1 : 1 : 1 \\
\hline
\phantom{$\frac{\hat I}{\hat I_1}$} 
{\bf 24} 
\phantom{$\frac{\hat I}{\hat I_1}$} 
& 
$-\frac{1}{2}$ : $-\frac{3}{2}$ : 1 \\
\hline
{\bf 75} & $-5$ : 3 : 1 \\
\hline
{\bf 200} & 10 : 2 : 1 \\
\hline
\end{tabular}
\caption{Ratios of gaugino masses for $F$-terms in representations of 
$SU(5)$, obtained in 
refs.~\cite{Ellis:1985jn},\cite{Anderson:1996bg}.\label{table:SU5}}
\end{table}
Only the $SU(5)$ singlet representation predicts universal gaugino masses. 
Depending on the model, one might suppose that one of the other three 
representations dominates the contribution to gaugino masses, or that two 
or more of the representations contribute in comparable amounts. The 
resulting phenomenology has been studied in many papers; for examples, see 
\cite{Anderson:1999uia}-\cite{Chattopadhyay:2009fr}.

The purpose of this paper is to extend these results to the case of 
unified groups $SO(10)$ and $E_6$, and all of their proper subgroups that 
embed $SU(3)_C \times SU(2)_L \times U(1)_Y$ in a way consistent with the 
Standard Model chiral fermion content. In each case, the $F$-terms can be 
classified by their transformation properties under both the full symmetry 
group and under subgroups that can be used to distinguish different 
Standard Model singlets. Then the object is to find the ratio of gaugino 
masses that can be produced by each distinct $F$-term representation. It 
is important to note that the subgroups are used only for distinguishing 
representations; they need not be the unbroken symmetry group for an 
effective theory at any scale. It is also possible that the full gauge 
symmetry is only a subgroup of the unified groups $SU(5)$, $SO(10)$ 
and $E_6$ in which the Standard Model gauge group can be embedded.
 
The method used to obtain the results below is as follows. For any given
unified symmetry group, one starts with a field transforming as the
symmetric product of the adjoint representation with itself,
$\Phi^{ab} = \langle F^{ab} \rangle/M_{\rm Planck}$. Under a gauge
transformation corresponding to a generator labeled by $c$,
the VEV transforms by an amount proportional to:
\beq
\delta_c \Phi^{ab} = (t^c)^{aa'} \Phi^{a'b} + (t^c)^{bb'} \Phi^{ab'} ,
\eeq
where the adjoint representation generators
are $(t^a)^{bc} = -i f^{abc}$, and $f^{abc}$ are the structure constants
of the Lie algebra. Since $\Phi^{ab}$ is required to be a Standard Model
singlet, one can require that $ \delta_{c} \Phi^{ab} = 0$ for
each of the 12 generators $c = 1,2,\ldots, 12$ of
$SU(3)_C \times SU(2)_L \times U(1)_Y$. This reduces $\Phi^{ab}$ 
from 300 independent entries to only 4 for $SU(5)$,  from
1035 entries to 9 for $SO(10)$, and from 3081
entries to 32 for $E_6$. This identifies the subspace of VEVs
that are Standard Model singlets. Now, to decompose $\Phi^{ab}$
into irreducible representations of the full symmetry group, 
one can use the quadratic Casimir operator:
\beq
C^{(ab),(a'b')} \Phi^{a'b'} \equiv \left [
(t^c t^c)^{aa'} \delta^{bb'} +
\delta^{aa'} (t^c t^c)^{bb'} + 2 (t^c)^{aa'} (t^c)^{bb'}\right ] \Phi^{a'b'}.
\label{eq:Casimir}
\eeq
The matrix $C$, realized for $SU(5)$, $SO(10)$, and $E_6$ respectively as 
a $4\times 4$, $9 \times 9$, and $32 \times 32$ dimensional matrix when 
restricted to the subspace of Standard Model singlets, has eigenvalues 
that are the quadratic Casimir invariants of the corresponding 
irreducible representations, and the corresponding eigenvectors are the 
Standard Model singlet members of the representations themselves. By 
restricting the implied sum over the index $c$ in eq.~(\ref{eq:Casimir}) 
to subsets of the full list of generators, one can likewise identify the 
quadratic Casimir invariant eigenvalues and thus the irreducible 
representations for any desired subgroup. It is an important and 
fortunate fact that the quadratic Casimir invariant eigenvalues are in 
one-to-one correspondence with the irreducible representations 
that appear
in all cases encountered below, so that ambiguities do not arise. One 
thus finds matrices $\Phi^{ab}$ that correspond to Standard Model 
singlets within any desired specific irreducible representations of the 
larger symmetry groups. These matrices $\Phi^{ab}$ are diagonal when the 
indices $a,b$ are restricted to the subspace of $SU(3)_C \times SU(2)_L 
\times U(1)_Y$ generators, with eigenvalues proportional to the 
corresponding gaugino masses $M_3$, $M_2$, and $M_1$.

I use the following notation for left-handed chiral fermion fields as
defined by their transformation properties under 
$SU(3)_C \times SU(2)_L \times U(1)_Y$:
\beq
&&Q \sim ({\bf 3}, {\bf 2}, 1/6),\qquad
\ubar \sim ({\bf \overline 3}, {\bf 1}, -2/3), \qquad
\dbar \sim ({\bf \overline 3}, {\bf 1}, 1/3),
\nonumber 
\\
&&L \sim ({\bf 1}, {\bf 2}, -1/2),\qquad
\ebar \sim ({\bf 1}, {\bf 1}, 1),\qquad
\nubar \sim ({\bf 1}, {\bf 1}, 0),\qquad
\nonumber
\\
&&H_u \sim ({\bf 1}, {\bf 2}, 1/2),\qquad
  H_d \sim ({\bf 1}, {\bf 2}, -1/2),\qquad
\nonumber
\\
&&  h \sim ({\bf  3}, {\bf 1}, -1/3), \qquad
  \overline{h} \sim ({\bf \overline 3}, {\bf 1}, 1/3), \qquad
  N \sim ({\bf 1}, {\bf 1}, 0).
\eeq
Here, $Q,\ubar,\dbar,L,\ebar$ are the notations for the Standard Model 
chiral fermions, and $\nubar$ for the additional Standard Model singlet 
found in the ${\bf 16}$ of $SO(10)$. The remaining fields $H_u$, $H_d$, 
$h$, $\overline h$, $N$ are the additional left-handed fermions found in 
the ${\bf 27}$ of $E_6$.

\section{Results for $SO(10)$ and its subgroups\label{sec:SO10}}
\setcounter{footnote}{1}
\setcounter{equation}{0}

In this section, I consider the unified group
$SO(10)$ \cite{SO10GUT}, for which the adjoint representation is the 
${\bf 45}$.
The possible $SO(10)$ irreducible representations for the $F$-term are:
\beq
({\bf 45} \times {\bf 45} )_S = {\bf 1} + {\bf 54} + {\bf 210} + {\bf 770}
.
\eeq
The ${\bf 1}$ and ${\bf 54}$ each contain exactly one Standard Model
singlet, but the ${\bf 210}$ contains three and the ${\bf 770}$ contains
four. To uniquely distinguish the possible $F$-terms,
one can additionally specify their transformation properties under a 
proper subgroup.
It turns out that choosing any one of the maximal proper subgroups is 
sufficient
to uniquely distinguish the possible $F$-terms in the $SO(10)$ case.

First, consider the ``normal" embedding of $SU(5) \times U(1) \subset 
SO(10)$,
with the fermions in a ${\bf 16}$ of $SO(10)$ transforming under $SU(5)$ as
$L, \dbar \sim {\bf \overline 5}$ and $Q, \ubar, \ebar \sim {\bf 10}$
and $\nubar \sim {\bf 1}$, as in the original Georgi-Glashow $SU(5)$ GUT 
model
\cite{SU5GUT}.
The ratios of gaugino masses obtained for $F$ terms in the various 
representations
of $SU(5) \subset SO(10)$ are shown in Table \ref{table:SO1051}. 
\begin{table}
\begin{tabular}[c]{|c|c|c|}
\hline
\phantom{X}$SO(10)$\phantom{X} & 
\phantom{X}$SU(5)$\phantom{X} &
$M_1$ : $M_2$ : $M_3$ \\
\hline\hline
{\bf 1} & {\bf 1} & 1 : 1 : 1 \\
\hline
\phantom{$\frac{\hat I}{\hat I_1}$} 
{\bf 54}
\phantom{$\frac{\hat I}{\hat I_1}$} 
 & {\bf 24} & 
$-\frac{1}{2}$ : $-\frac{3}{2}$ : 1 
\\
\hline
{\bf 210}  & 
{\bf 1} & 1 : 1 : 1 \\
    & {\bf 24} & 
$-\frac{1}{2}$ : $-\frac{3}{2}$ : 1 
\\
    & {\bf 75} & $-5$ : 3 : 1 \\
\hline
{\bf 770}  & {\bf 1} & 1 : 1 : 1 \\
    & {\bf 24} & 
$-\frac{1}{2}$ : $-\frac{3}{2}$ : 1 
\\
    & {\bf 75} & $-5$ : 3 : 1 \\
    & {\bf 200} & 10 : 2 : 1 \\
\hline
\end{tabular}
\caption{Ratios of gaugino masses for $F$-terms in representations of
$SU(5) \subset SO(10)$, with the normal (non-flipped) 
embedding.\label{table:SO1051}}
\end{table}
Note that these results merely agree with those already listed for
$SU(5)$ in Table \ref{table:SU5}; this follows from the fact that
the Standard Model generators are entirely embedded within the 
$SU(5)$ simple subgroup. In several similar situations below, 
where the gaugino mass ratios follow trivially from the results already 
obtained from a subgroup, I will just note this fact rather than record 
the results in tabular form.

The ``flipped" embedding \cite{flippedSU5} of $SU(5)' \times U(1) \subset 
SO(10)$ yields different results, because the $U(1)_Y$ generator
is a linear combination of the $U(1)$ generator 
inside $SU(5)'$ that commutes with $SU(3)_C \times SU(2)_L$ and the 
$U(1)$ generator outside of $SU(5)'$. In this case, the fermions
in a ${\bf 16}$ of $SO(10)$ transform under $SU(5)' \times U(1)$ as
$L, \ubar \sim ({\bf \overline 5}, 3)$ and $Q, \dbar, \nubar \sim
({\bf 10}, -1)$ and $\ebar \sim ({\bf 1}, -5)$. The results for
gaugino mass ratios are shown in Table \ref{table:SO1051flipped}.
\begin{table}
\begin{tabular}[c]{|c|c|c|}
\hline
\phantom{X}$SO(10)$\phantom{X} & 
\phantom{X}$[SU(5)'\times U(1)]_{\rm flipped}$\phantom{X} &
$M_1$ : $M_2$ : $M_3$ \\
\hline\hline
{\bf 1} & ({\bf 1}, 0) & 1 : 1 : 1 \\
\hline
{\bf 54} & 
\phantom{$\frac{\hat I}{\hat I_1}$} 
({\bf 24}, 0) 
\phantom{$\frac{\hat I}{\hat I_1}$} 
& \phantom{x}
$-\frac{1}{2}$ : $-\frac{3}{2}$ : 1 \phantom{x} \\
\hline
{\bf 210}  & ({\bf 1}, 0) & $-\frac{19}{5}$ : 1 : 1 \\
    & ({\bf 24}, 0) & $\frac{7}{10}$ : $-\frac{3}{2}$ : 1 \\
\phantom{$\frac{\hat I}{\hat I_1}$} 
    & ({\bf 75}, 0) & $-\frac{1}{5}$ : 3 : 1 \\
\hline
{\bf 770}  & ({\bf 1}, 0) & $\frac{77}{5}$ : 1 : 1 \\
    & ({\bf 24}, 0) & $-\frac{101}{10}$ : $-\frac{3}{2}$ : 1 \\
    & ({\bf 75}, 0) & $-\frac{1}{5}$ : 3 : 1 \\
\phantom{$\frac{\hat I}{\hat I_1}$} 
    & ({\bf 200}, 0) & $\frac{2}{5}$ : 2 : 1 \\
\hline
\end{tabular}
\caption{Ratios of gaugino masses for $F$-terms in representations of 
flipped 
$SU(5)' \times U(1) \subset SO(10)$.\label{table:SO1051flipped}}
\end{table}
Note that the results for the ratio $M_2/M_3$ are the
same as for the normal $SU(5)$ embedding, but the $M_1/M_2$
and $M_1/M_3$ ratios are different.

In some models, it is more useful to distinguish the possible 
$F$-terms by their transformations under the maximal proper subgroup 
$SU(4) \times SU(2)_L \times SU(2)_R$, where the $SU(4)$ is the 
Pati-Salam unification \cite{PatiSalam} of color and $B-L$. Here the 
fermions in a ${\bf 16}$ of $SO(10)$ transform under $SU(4) \times 
SU(2)_L \times SU(2)_R$ as $Q,L \sim ({\bf 4}, {\bf 2}, {\bf 1})$ and 
$\ubar, \dbar, \nubar, \ebar \sim ({\bf \overline 4}, {\bf 1}, {\bf 2})$. 
The results for gaugino mass ratios for $F$-terms classified by 
representations within this subgroup are shown in Table \ref{table:SO10422}.
\begin{table}
\begin{tabular}[c]{|c|c|c|}
\hline
\phantom{X}$SO(10)$\phantom{X} & 
\phantom{X}$SU(4) 
\times SU(2)_R$\phantom{X} &
$M_1$ : $M_2$ : $M_3$ \\
\hline\hline
{\bf 1} & ({\bf 1},  {\bf 1}) & 1 : 1 : 1 \\
\hline
{\bf 54} & 
\phantom{$\frac{\hat I}{\hat I_1}$} 
({\bf 1}, {\bf 1}) 
\phantom{$\frac{\hat I}{\hat I_1}$} 
& \phantom{x}
$-\frac{1}{2}$ : $-\frac{3}{2}$ : 1 \phantom{x} \\
\hline
{\bf 210}  & ({\bf 1},  {\bf 1}) & $-\frac{3}{5}$ : 1 : 0 \\
    & ({\bf 15},  {\bf 1}) & $-\frac{4}{5}$ : 0 : 1 \\
    & ({\bf 15},  {\bf 3}) & 1 : 0 : 0 \\
\hline
{\bf 770}  & ({\bf 1},  {\bf 1}) & $\frac{19}{10}$ : $\frac{5}{2}$ : 1 \\
    & ({\bf 1},  {\bf 5}) & 1 : 0 : 0 \\
    & ({\bf 15},  {\bf 3}) & 1 : 0 : 0 \\
\phantom{$\frac{\hat I}{\hat I_1}$} 
    & ({\bf 84},  {\bf 1}) & $\frac{32}{5}$ : 0 : 1 \\
\hline
\end{tabular}
\caption{Ratios of gaugino masses for $F$-terms in representations of $SU(4) \times SU(2)_L \times SU(2)_R
\subset SO(10)$.\label{table:SO10422}}
\end{table}

The above is a complete description of the situation for $SO(10)$, since 
the three maximal proper subgroup embeddings listed above are the only 
ones compatible with the Standard Model fermion assignments, and the 
resulting gaugino mass ratios are completely determined in each case. 
Classification of $F$-terms in representations of smaller subgroups of 
$SO(10)$ are obtained simply as special cases of the above results.

Results equivalent to those of lines 2, 3, and 6 of Table IV in the 
present paper have already been given (presented in the $SU(4) \times 
SU(2)_L \times SU(2)_R$ basis for gaugino masses rather than the MSSM 
basis $SU(3)_C \times SU(2)_L \times U(1)_Y$ basis $M_3, M_2, M_1$ 
used here) in 
ref.~\cite{Chakrabortty:2008zk}, see Table 3 and eq.~(31).

The results given in this section do not agree with those 
reported in ref.~\cite{Chamoun:2001in}, which have been used as the basis 
for several studies in the literature. In ref.~\cite{Chamoun:2001in}, it 
was claimed that an $F$-term in the ${\bf 54}$ representation with a 
breaking chain including $SU(4) \times SU(2)_L \times SU(2)_R$ would give 
$M_1 : M_2 : M_3 = -1 : -3/2 : 1$. However, there is only one singlet 
within the ${\bf 54}$ of $SU(10)$ that can give masses to the MSSM 
gauginos, and that singlet also resides within the ${\bf 24}$ of the
Georgi-Glashow $SU(5)$. 
Therefore, as shown in Tables \ref{table:SU5}-\ref{table:SO10422} 
above, it must give gaugino mass ratios that agree with the $SU(5)$ 
${\bf 24}$ case, regardless of the VEV classification scheme or the 
breaking pattern. Ref.~\cite{Chamoun:2001in} also gives a result for breaking 
through $SU(2) \times SO(7)$, but this is impossible to reconcile with 
the chiral fermion assignments of the Standard Model, 
since $SO(7)$ 
spinor representations are real. Finally, it is claimed 
in ref.~\cite{Chamoun:2001in} that for the ${\bf 210}$ of $SO(10)$ with 
the breaking chain through flipped $SU(5)' \times U(1)$, the gaugino mass 
ratios are $M_1 : M_2 : M_3 = -96/25 : 1 : 1$, but 
this should actually be $-19/5 : 1 : 1$, as seen in the third 
line of Table \ref{table:SO1051flipped} above.

\section{Results for $E_6$ and its subgroups\label{sec:E6}}
\setcounter{footnote}{1}
\setcounter{equation}{0}

In this section, I consider the unified group
$E_6$ \cite{E6GUT}, 
for which the adjoint representation is the ${\bf 78}$.
The possible $E_6$ irreducible representations for the $F$-term are:
\beq
({\bf 78} \times {\bf 78} )_S = {\bf 1} + {\bf 650} + {\bf 2430} .
\eeq
The ${\bf 1}$, the ${\bf 650}$, and the ${\bf 2430}$ 
contain respectively 1, 11, and 20 Standard Model singlets.
To uniquely distinguish the possible $F$-terms,
one can specify their transformation properties under various subgroups,
as in the $SO(10)$ case. 
Many of these possible $F$-term representations are not capable of
giving any masses to MSSM gauginos; 
they can be recognized as those that 
are charged under any $U(1)$ subgroup of $E_6$, 
or transform non-trivially under
the subgroup $SU(2)_X$ under which $(L, H_d)$ and $(\dbar, \overline h)$
and $(\nubar, N)$ are doublets.
Representations that do not contribute to MSSM gaugino masses
are omitted from the tables below. 

First, consider the ``normal" embeddings of $SO(10) \subset E_6$,
with the chiral fields transforming as  
$L, \dbar, Q, \ubar, \ebar, \nubar \sim {\bf 16}$ and 
$H_d, \overline h, H_u, h \sim {\bf 10}$ and
$N \sim {\bf 1}$ under $SO(10)$. Then there are three possible
distinct embeddings of the chiral fields into maximal proper subgroups
of $SO(10)$, corresponding to the cases discussed in the previous section:
normal $SU(5)$, flipped $SU(5)' \times U(1)$, and
$SU(4) \times SU(2)_L \times SU(2)_R$. To find the ratios of
gaugino masses, it is only necessary to know how the 
relevant 
representations within $E_6$ break 
down into $SO(10)$ representations:
\beq
{\bf 1} &\rightarrow & {\bf 1},
\\
{\bf 650} &\rightarrow & {\bf 1} + {\bf 54} + {\bf 210} + \left (
{\bf 10} + {\bf 10} + {\bf 16} + {\bf \overline{16}} + {\bf 45}  + 
{\bf 144} + {\bf \overline{144}} \right )
,
\\
{\bf 2430} &\rightarrow & {\bf 1} + {\bf 210} + {\bf 770} 
+ \left ( {\bf 16} + {\bf \overline{16}} + {\bf 45} + 
{\bf 126} + {\bf \overline{126}} + {\bf 560} + {\bf \overline{560}} \right ).
\eeq
The representations enclosed in parentheses cannot contribute to MSSM gaugino
masses in this embedding, even though all except the ${\bf 10}$'s contain
at least one Standard Model singlet. 
For the other representations of $SO(10)$, namely the ${\bf 1}$, 
${\bf 54}$, ${\bf 210}$, and ${\bf 770}$, the corresponding 
gaugino mass ratio contributions simply 
follow immediately from the results of Tables
\ref{table:SO1051}, \ref{table:SO1051flipped}, and \ref{table:SO10422},
in a way just analogous to how Table \ref{table:SO1051} follows from
Table \ref{table:SU5}.

Next, consider the ``flipped $SO(10)$" embedding 
\cite{flippedSO10} within $E_6$,
for which the chiral fields transform under $SO(10)' \times U(1)$
as $Q, \overline h, N, H_u, \dbar, \nubar \sim ({\bf 16},1)$ 
and $h, L, \ubar, H_d \sim ({\bf 10}, -2)$ and
$\bar e \sim ({\bf 1}, 4)$. This does not reduce to the results
of the preceding section, because the Standard Model $U(1)_Y$ generator
is not contained within the $SO(10)'$ subgroup. 
There are two distinct subcases,
in which the $F$-terms are classified by their representations
under a maximal proper subgroup of $SO(10)'$, either 
$SU(5)'' \times U(1)'$ or $SU(4)' \times SU(2)_L \times SU(2)_X$.

In the first subcase of flipped $SO(10)$, 
the chiral fields transform under $SU(5)'' \times U(1)'$ as
$Q, \overline h, N \sim ({\bf 10}, -1)$ and 
$H_u, \dbar \sim ({\bf \overline 5}, 3)$ and
$h, L \sim ({\bf 5}, 2)$ and $\ubar, H_d \sim ({\bf \overline 5}, -2)$ and
$\ebar \sim ({\bf 1}, -5)$. 
The results for the gaugino mass ratios for $F$-term representations
classified by this subgroup are shown in Table \ref{table:E610flipped51}.
\begin{table}
\begin{tabular}[c]{|c|c|c|c|}
\hline
\phantom{X}$E_6$\phantom{X} & \phantom{X}$[SO(10)' \times 
U(1)]_{\rm flipped} $\phantom{X} &
\phantom{X}$SU(5)''$\phantom{X} &
\phantom{X}$M_1$ : $M_2$ : $M_3$\phantom{X}\\
\hline\hline
{\bf 1} & ({\bf 1}, 0) & {\bf 1} & 1 : 1 : 1 \\
\hline
{\bf 650}  & 
\phantom{$\frac{\hat I}{\hat I_1}$} 
({\bf 1}, 0) 
\phantom{$\frac{\hat I}{\hat I_1}$} 
& {\bf 1} & $-\frac{22}{5}$ : $1$ : $1$
\\
\cline{2-4}
& ({\bf 45}, 0 ) &{\bf 1} & $1$ : $0$ : $0$
\\
& & {\bf 24} & $1$ : $0$ : $0$
\\
\cline{2-4}
& 
\phantom{$\frac{\hat I}{\hat I_1}$} 
({\bf 54}, 0) 
\phantom{$\frac{\hat I}{\hat I_1}$} 
& {\bf 24} & $\frac{1}{10}$ : $-\frac{3}{2}$ : $1$
\\
\cline{2-4}
& ({\bf 210}, 0) & {\bf 1} & $-\frac{1}{5}$ : $1$ : $1$
\\
& & {\bf 24} & $-\frac{1}{5}$ : $-\frac{3}{2}$ : $1$
\\
& 
\phantom{$\frac{\hat I}{\hat I_1}$} 
& {\bf 75} & $-\frac{1}{5}$ : $3$ : $1$
\\
\hline
{\bf 2430} & 
\phantom{$\frac{\hat I}{\hat I_1}$} 
({\bf 1}, 0) 
\phantom{$\frac{\hat I}{\hat I_1}$} 
& {\bf 1} & $\frac{122}{5}$ : $1$ : $1$
\\
\cline{2-4}
& ({\bf 45}, 0) & {\bf 1} & $1$ : $0$ : $0$
\\
& & {\bf 24} & $1$ : $0$ : $0$
\\
\cline{2-4}
& ({\bf 210}, 0) & {\bf 1} & $-\frac{1}{5}$ : $1$ : $1$
\\
& & {\bf 24} & $-\frac{1}{5}$ : $-\frac{3}{2}$ : $1$
\\
& 
\phantom{$\frac{\hat I}{\hat I_1}$} 
& {\bf 75} & $-\frac{1}{5}$ : $3$ : $1$
\\
\cline{2-4}
& ({\bf 770}, 0) & {\bf 1} & $1$ : $1$ : $1$
\\
& & {\bf 24} & $\frac{5}{2}$ : $-\frac{3}{2}$ : $1$
\\
& & {\bf 75} & $-\frac{1}{5}$ : $3$ : $1$
\\
& 
\phantom{$\frac{\hat I}{\hat I_1}$} 
& {\bf 200} & $\frac{2}{5}$ : $2$ : $1$
\\
\hline
\end{tabular}
\caption{Ratios of gaugino masses for $F$-terms in representations of
 $SU(5)''\times U(1)' \times U(1) 
\subset [SO(10)' \times U(1)]_{\rm flipped}
\subset E_6$.\label{table:E610flipped51}}
\end{table}

In the second subcase of flipped $SO(10)$, 
the chiral fields transform under $SU(4)' \times SU(2)_L \times SU(2)_X
\subset SO(10)$ as
$Q, H_u \sim ({\bf 4}, {\bf 2}, {\bf 1})$ and 
$\dbar, \nubar, \overline h, N \sim 
({\bf \overline 4}, {\bf 1}, {\bf 2})$ and
$\ubar, h \sim ({\bf 6}, {\bf 1}, {\bf 1})$ and 
$L, H_d \sim ({\bf 1}, {\bf 2}, {\bf 2})$ and
$\ebar \sim ({\bf 1}, {\bf 1}, {\bf 1})$. 
The results for the gaugino mass ratios for $F$-term representations
classified by this subgroup are shown in Table \ref{table:E610flipped422}.
\begin{table}
\begin{tabular}[c]{|c|c|c|c|}
\hline
\phantom{X}$E_6$\phantom{X} & \phantom{X}$[SO(10)' \times 
U(1)]_{\rm flipped} $\phantom{X} &
\phantom{X}$SU(4)' $\phantom{X} &
\phantom{X}$M_1$ : $M_2$ : $M_3$\phantom{X}\\
\hline\hline
{\bf 1} & ({\bf 1}, 0) & {\bf 1} & 1 : 1 : 1 \\
\hline
{\bf 650}  & 
\phantom{$\frac{\hat I}{\hat I_1}$} 
({\bf 1}, 0) 
\phantom{$\frac{\hat I}{\hat I_1}$} 
& {\bf 1} & $-\frac{22}{5}$ : $1$ : $1$
\\
\cline{2-4}
& ({\bf 45}, 0 ) & {\bf 15} & $1$ : $0$ : $0$
\\
\cline{2-4}
& 
\phantom{$\frac{\hat I}{\hat I_1}$} 
({\bf 54}, 0) 
\phantom{$\frac{\hat I}{\hat I_1}$} 
& {\bf 1} & $\frac{1}{10}$ : $-\frac{3}{2}$ : $1$
\\
\cline{2-4}
& ({\bf 210}, 0) & {\bf 1} & $0$ : $1$ : $0$
\\
&\phantom{$\frac{\hat I}{\hat I_1}$} 
 & {\bf 15} & $-\frac{1}{5}$ : $0$ : $1$
\\
\hline
{\bf 2430} & 
\phantom{$\frac{\hat I}{\hat I_1}$} 
({\bf 1}, 0) 
\phantom{$\frac{\hat I}{\hat I_1}$} 
& {\bf 1} & $\frac{122}{5}$ : $1$ : $1$
\\
\cline{2-4}
& ({\bf 45}, 0) & {\bf 15} & $1$ : $0$ : $0$
\\
\cline{2-4}
& ({\bf 210}, 0) & {\bf 1} & $0$ : $1$ : $0$
\\
& 
\phantom{$\frac{\hat I}{\hat I_1}$} 
& {\bf 15} & $-\frac{1}{5}$ : $0$ : $1$
\\
\cline{2-4}
& ({\bf 770}, 0) & {\bf 1} & $\frac{1}{10}$ : 
$\frac{5}{2}$ : $1$
\\
& 
\phantom{$\frac{\hat I}{\hat I_1}$} 
& {\bf 84} & $\frac{8}{5}$ : $0$ : $1$
\\
\hline
\end{tabular}
\caption{Ratios of gaugino masses for $F$-terms in representations of $SU(4)'\times SU(2)_L \times SU(2)_X \times 
U(1) \subset [SO(10)' \times U(1)]_{\rm flipped}
\subset E_6$. These $F$ terms are all singlets under $SU(2)_X$.
\label{table:E610flipped422}}
\end{table}

Another useful way to classify the possible $F$-terms is through
representations of the trinification \cite{trinification} subgroup of $E_6$,
$SU(3)_C \times SU(3)_L \times SU(3)_R$. The chiral
fields transform under this group as
$Q, h \sim ({\bf 3}, {\bf 3}, {\bf 1})$ and
$\dbar, \ubar, \overline h \sim ({\bf \overline 3}, {\bf 1}, {\bf 
\overline 3})$ and
$L, H_u, H_d, \ebar, \nubar, N \sim ({\bf 1}, {\bf \overline 3}, {\bf 3})$.
The results for the ratios of gaugino masses are given in Table 
\ref{table:E6333}.
\begin{table}
\begin{tabular}[c]{|c|c|c|}
\hline
\phantom{X}$E_6$\phantom{X} & 
\phantom{X}$SU(3)_L  
\times SU(3)_R$\phantom{X} & \ \phantom{x}
$M_1$ : $M_2$ : $M_3$ \phantom{x} \\
\hline\hline
{\bf 1} & ({\bf 1}, {\bf 1}) & 1 : 1 : 1 \\
\hline
{\bf 650}  & 
    \phantom{${}_1$}({\bf 1}, {\bf 1})$_1$ & $-\frac{3}{5}$ : $1$ : $0$
\\
  & \phantom{${}_2$}({\bf 1}, {\bf 1})$_2$ &  $-\frac{4}{5}$ : $0$ : $1$
\\
    & ({\bf 1}, {\bf 8}) & $1$ : 0 : 0 \\
    & ({\bf 8}, {\bf 1}) & $-\frac{1}{5}$ : 1 : 0 \\
    & ({\bf 8}, {\bf 8}) & 1 : 0 : 0 \\
\hline
{\bf 2430}  & ({\bf 1}, {\bf 1}) & 1 : 1 : 1 \\
    & ({\bf 1}, {\bf 8}) & 1 : 0 : 0 \\
    & ({\bf 8}, {\bf 1}) & $-\frac{1}{5}$ : 1 : 0 \\
    & ({\bf 8}, {\bf 8}) & 1 : 0 : 0 \\
    & ({\bf 1}, {\bf 27}) & 1 : 0 : 0 \\
\phantom{$\frac{\hat I}{\hat I_1}$} 
    & ({\bf 27}, {\bf 1}) & $\frac{9}{5}$ : 1 : 0 \\
\hline
\end{tabular}
\caption{Ratios of gaugino masses for $F$-terms in representations of 
the trinification subgroup
$SU(3)_C \times SU(3)_L \times SU(3)_R \subset E_6$.\label{table:E6333}}
\end{table}
In this classification scheme, there are two distinct singlets of
$SU(3)_C \times SU(3)_L \times SU(3)_R$ within the ${\bf 650}$ of
$E_6$, which cannot be distinguished by their transformations
under this subgroup. The results given in the table represent
an arbitrary choice of basis for these two representations, denoted
by subscripts $1$ and $2$. The results of lines 2, 3, and 7 of Table
\ref{table:E6333} are equivalent to those obtained previously in
\cite{Chakrabortty:2008zk} (see Table 4 and eq.~(32); this reference
presents the gaugino mass contributions in the $SU(3)_C \times SU(3)_L 
\times 
SU(3)_R$ basis rather than the MSSM $SU(3)_C \times SU(2)_L \times
U(1)_Y$ basis $M_3$, $M_2$, $M_1$ used here).

Finally, one can classify $F$-term representations by their transformations
under the maximal proper subgroup $SU(6) \times SU(2) \subset E_6$.
There are three distinct possible embeddings of the chiral fields within
this group, which can be treated as subcases. 

In the first subcase, one classifies possible $F$-term
representations by $SU(6) \times SU(2)_X$, where the chiral fields
are assigned to representations 
$\dbar, \overline h, \nubar, N, L, H_d
\sim ({\bf \overline 6}, {\bf 2})$ and 
$Q, H_u, h, \ubar, \ebar \sim 
({\bf 15}, {\bf 1})$. To uniquely distinguish the
possible $F$-terms, it is sufficient to 
also give the representation
under the $SU(3)_L$ subgroup of $SU(6)$ that commutes with $SU(3)_C$.
This $SU(3)_L$ contains the weak isospin group $SU(2)_L$.  
The results for the ratios of gaugino masses are given in Table
\ref{table:E662X}.
\begin{table}
\begin{tabular}[c]{|c|c|c|c|}
\hline
\phantom{X}$E_6$\phantom{X} & 
\phantom{X}$SU(6)\times SU(2)_X$\phantom{X} & 
\phantom{x}$SU(3)_L$\phantom{x}
& \phantom{x}$M_1$ : $M_2$ : $M_3$ \phantom{x}\\
\hline\hline
{\bf 1} & ({\bf 1}, {\bf 1}) & {\bf 1} & 1 : 1 : 1 \\
\hline
{\bf 650}  & ({\bf 1}, {\bf 1}) & {\bf 1} & $1$ : $1$ : $1$
\\
\cline{2-4}
& ({\bf 35}, {\bf 1}) & \phantom{$\frac{\hat I}{\hat I_1}$} 
{\bf 1} 
\phantom{$\frac{\hat I}{\hat I_1}$} 
& $-\frac{1}{5}$ : $-1$ : $1$
\\
& 
\phantom{$\frac{\hat I}{\hat I_1}$} 
& {\bf 8} & $\frac{3}{5}$ : $1$ : $0$
\\
\cline{2-4}
& ({\bf 189}, {\bf 1}) & {\bf 1} & $-3$ : $1$ : $1$
\\
& & {\bf 8} & $-1$ : $1$ : $0$
\\
\hline
{\bf 2430} & ({\bf 1}, {\bf 1}) & {\bf 1} & $1$ : $1$ : $1$
\\
\cline{2-4}
& ({\bf 189}, {\bf 1}) & {\bf 1} & $-3$ : $1$ : $1$
\\
& & {\bf 8} & $-1$ : $1$ : $0$
\\
\cline{2-4}
& ({\bf 405}, {\bf 1}) & \phantom{$\frac{\hat I}{\hat I_1}$} 
{\bf 1} 
\phantom{$\frac{\hat I}{\hat I_1}$} 
& $\frac{33}{5}$ : $1$ : $1$
\\
& \phantom{$\frac{\hat I}{\hat I_1}$} 
& {\bf 8} & $\frac{19}{5}$ : $1$ : $0$
\\
& \phantom{$\frac{\hat I}{\hat I_1}$} 
& {\bf 27} & $\frac{9}{5}$ : $1$ : $0$
\\
\hline
\end{tabular}
\caption{Ratios of gaugino masses for $F$-terms in representations of
$SU(3)_C \times SU(3)_L \times U(1) 
\times SU(2)_X \subset SU(6)\times SU(2)_X 
\subset E_6$.\label{table:E662X}}
\end{table}
As noted earlier, the only representations that can give
non-zero gaugino masses are singlets under $SU(2)_X$.

In the second subcase, one classifies $F$-terms by their
representations under $SU(6)' \times SU(2)_R$, where the chiral fields
are assigned to representations $\ubar, \dbar, \ebar, \nubar, H_u, H_d
\sim ({\bf \overline 6}, {\bf 2})$ and $Q, L, h, \overline h, N \sim 
({\bf 15}, {\bf 1})$. To uniquely distinguish the
different possible $F$-terms, it is again sufficient to 
also give the representation
under the $SU(3)_L$ subgroup of $SU(6)'$ that commutes with $SU(3)_C$.  
The results for the ratios of gaugino masses are given in Table
\ref{table:E662R}.
\begin{table}
\begin{tabular}[c]{|c|c|c|c|}
\hline
\phantom{X}$E_6$\phantom{X} & 
\phantom{X}$SU(6)'\times SU(2)_R$\phantom{X} & \phantom{x}$SU(3)_L$\phantom{x}
& \phantom{x} $M_1$ : $M_2$ : $M_3$ \phantom{x}\\
\hline\hline
{\bf 1} & ({\bf 1}, {\bf 1}) & {\bf 1} & 1 : 1 : 1 \\
\hline
{\bf 650}  & 
\phantom{$\frac{\hat I}{\hat I_1}$} 
({\bf 1}, {\bf 1}) 
\phantom{$\frac{\hat I}{\hat I_1}$} 
& {\bf 1} & $-\frac{13}{5}$ : $1$ : $1$
\\
\cline{2-4}
& ({\bf 35}, {\bf 1}) & {\bf 1} & $-\frac{1}{5}$ : $-1$ : $1$
\\
& 
\phantom{$\frac{\hat I}{\hat I_1}$} 
& {\bf 8} & $-\frac{3}{5}$ : $1$ : $0$
\\
\cline{2-4}
& ({\bf 35}, {\bf 3}) & {\bf 1} & $1$ : $0$ : $0$
\\
& & {\bf 8} & $1$ : $0$ : $0$
\\
\cline{2-4}
& ({\bf 189}, {\bf 1}) & {\bf 1} & $-\frac{3}{5}$ : $1$ : $1$
\\
& \phantom{$\frac{\hat I}{\hat I_1}$} 
& {\bf 8} & $\frac{1}{5}$ : $1$ : $0$
\\
\hline
{\bf 2430} & 
\phantom{$\frac{\hat I}{\hat I_1}$} 
({\bf 1}, {\bf 1}) 
\phantom{$\frac{\hat I}{\hat I_1}$} 
& {\bf 1} & $\frac{41}{15}$ : $1$ : $1$
\\
\cline{2-4}
& ({\bf 1}, {\bf 5}) & {\bf 1} & $1$ : $0$ : $0$
\\
\cline{2-4}
& ({\bf 35}, {\bf 3}) & {\bf 1} & $1$ : $0$ : $0$
\\
& & {\bf 8} & $1$ : $0$ : $0$
\\
\cline{2-4}
& ({\bf 189}, {\bf 1}) & 
\phantom{$\frac{\hat I}{\hat I_1}$} 
{\bf 1} 
\phantom{$\frac{\hat I}{\hat I_1}$} 
& $-\frac{3}{5}$ : $1$ : $1$
\\
& \phantom{$\frac{\hat I}{\hat I_1}$} 
 & {\bf 8} & $\frac{1}{5}$ : $1$ : $0$
\\
\cline{2-4}
& ({\bf 405}, {\bf 1}) & 
\phantom{$\frac{\hat I}{\hat I_1}$} 
{\bf 1} 
\phantom{$\frac{\hat I}{\hat I_1}$} 
& $\frac{9}{5}$ : $1$ : $1$
\\
&\phantom{$\frac{\hat I}{\hat I_1}$} 
 & {\bf 8} & $-\frac{11}{5}$ : $1$ : $0$
\\
& \phantom{$\frac{\hat I}{\hat I_1}$} 
& {\bf 27} & $\frac{9}{5}$ : $1$ : $0$
\\
\hline
\end{tabular}
\caption{Ratios of gaugino masses for $F$-terms in representations of
$SU(3)_C \times SU(3)_L \times U(1) 
\times SU(2)_R \subset SU(6)'\times SU(2)_R 
\subset E_6$.\label{table:E662R}}
\end{table}

The last subcase identifies the $SU(2)$ with the weak isospin of the
Standard Model, so that the chiral fields transform under 
$SU(6)'' \times SU(2)_L$
as $Q, L, H_u, H_d \sim ({\bf 6}, {\bf 2})$ and $\ubar,\dbar,\nubar,\ebar,h,\overline h, N
\sim ({\bf \overline{15}}, {\bf 1})$. To uniquely distinguish the
different possible $F$-terms in this case, it is sufficient to 
also give the representation
under the $SU(3)_R$ subgroup of $SU(6)''$ that commutes with $SU(3)_C$.
The results for the ratios of gaugino masses are given in Table
\ref{table:E662L}.
\begin{table}
\begin{tabular}[c]{|c|c|c|c|}
\hline
\phantom{X}$E_6$\phantom{X} & 
\phantom{X}$SU(6)''\times SU(2)_L$\phantom{X} & \phantom{x} $SU(3)_R$\phantom{x} 
& \phantom{x} $M_1$ : $M_2$ : $M_3$ \phantom{x} \\
\hline\hline
{\bf 1} & ({\bf 1}, {\bf 1}) & {\bf 1} & 1 : 1 : 1 \\
\hline
{\bf 650}  & ({\bf 1}, {\bf 1}) & {\bf 1} & $1$ : $-5$ : $1$
\\
\cline{2-4}
& ({\bf 35}, {\bf 1}) & 
\phantom{$\frac{\hat I}{\hat I_1}$} 
{\bf 1} 
\phantom{$\frac{\hat I}{\hat I_1}$} 
& $-\frac{4}{5}$ : $0$ : $1$
\\
\cline{2-4}
& ({\bf 189}, {\bf 1}) & {\bf 1} & $0$ : $0$ : $1$
\\
& & {\bf 8} & $1$ : $0$ : $0$
\\
\hline
{\bf 2430} & ({\bf 1}, {\bf 1}) & 
\phantom{$\frac{\hat I}{\hat I_1}$} 
{\bf 1} 
\phantom{$\frac{\hat I}{\hat I_1}$} 
& $1$ : $\frac{35}{9}$ : $1$
\\
\cline{2-4}
& ({\bf 189}, {\bf 1}) & {\bf 1} & $0$ : $0$ : $1$
\\
& & {\bf 8} & $1$ : $0$ : $0$
\\
\cline{2-4}
& ({\bf 405}, {\bf 1}) & 
\phantom{$\frac{\hat I}{\hat I_1}$} 
{\bf 1} 
\phantom{$\frac{\hat I}{\hat I_1}$} 
& $\frac{12}{5}$ : $0$ : $1$
\\
& & {\bf 8} & $1$ : $0$ : $0$
\\
& & {\bf 27} & $1$ : $0$ : $0$
\\
\hline
\end{tabular}
\caption{Ratios of gaugino masses for $F$-terms in representations of
$SU(3)_C \times SU(3)_R \times U(1) 
\times SU(2)_L \subset SU(6)''\times SU(2)_L 
\subset E_6$.\label{table:E662L}}
\end{table}

This concludes the discussion of $E_6$, since these are the only maximal 
proper subgroup embeddings
that are consistent with the Standard Model fermion content. 
Here, we do not count as distinct any embeddings that differ from the 
above ones by exchanging the identities of ($H_d$, $L$) or ($\nubar, N$) or 
($\dbar, \overline h$), since these pairs of fields 
have the same $SU(3)_C \times SU(2)_L 
\times U(1)_Y$ quantum numbers. Such embeddings will give the same
results for gaugino mass ratios as the ones listed above.

\clearpage
\section{Concluding remarks\label{sec:conclusion}}
\setcounter{footnote}{1}
\setcounter{equation}{0}

\baselineskip=13.975pt

In this paper, I have derived the ratios of gaugino masses that follow 
from non-singlet $F$-term VEVs in various representations of the unified 
groups $SO(10)$ and $E_6$, 
and their subgroups, in which the Standard Model gauge group can be 
embedded. One interesting facet of these results is that in models with 
$SO(10)$ symmetry, if one views the most likely 
large deviation from universality as coming from the smallest 
non-singlet representation, 
namely the ${\bf 54}$, then the contribution to the gaugino mass ratios 
is necessarily the same as that found from the ${\bf 24}$ of $SU(5)$,
which is distinguished by the same criterion.
However, clearly any desired deviation from universality 
can be achieved by
taking an $F$-term VEV in an appropriate linear combination of 
representations, in a variety of ways.
 
Some of the results found here (lines 2, 3, and 6 of Table 
\ref{table:SO10422} and lines 2, 3, and 7 of Table \ref{table:E6333}) 
agree with those obtained earlier and presented in a different form in 
ref.~\cite{Chakrabortty:2008zk}. However, my results disagree with those 
found in ref.~\cite{Chamoun:2001in}.

In this paper, I have not relied on any particular symmetry breaking 
pattern, instead using the representations of unified and partially 
unified symmetry groups only as a classification scheme for the possible 
$F$ terms. How these results might be realized in particular models is a 
separate and detailed dynamical question. It should be noted that in most 
viable supersymmetric GUT models, the gauge coupling above the unification 
scale probably is quickly driven so large by renormalization group running 
as to render perturbative analyses problematic. Threshold corrections can 
also be quite significant \cite{SU5thresholds}, and in general they will 
have a different structure than the tree-level gaugino mass ratios given 
here. There are many other ways that gaugino mass non-universality can be 
realized in model building. However, if the deviation from universality 
observed in gaugino masses is sufficiently dramatic, and can be correlated 
with other observed features of the MSSM, then the ratios of gaugino 
masses following from non-singlet $F$-terms may yield an important insight 
into the structure of supersymmetry breaking.

\textit{Acknowledgments:} This work was supported in part by National 
Science Foundation grant PHY-0757325.



\begin{thebibliography}{90}
\baselineskip=12.6pt

\bibitem{MSSMreviews} For reviews, see for example
M.~Drees, R.~Godbole and P.~Roy,
  ``Theory and phenomenology of sparticles: An account of 
  four-dimensional N=1
  supersymmetry in high energy physics,''
{\it  World Scientific (2004)};
  H.~Baer and X.~Tata,
  ``Weak scale supersymmetry: From superfields to scattering events,''
{\it  Cambridge University Press (2006)};
  S.P.~Martin,
  ``A supersymmetry primer,''
  [hep-ph/9709356] (version 5, December 2008).

\bibitem{KaneKing}
  G.L.~Kane and S.F.~King,
  Phys.\ Lett.\  B {\bf 451}, 113 (1999)
  [hep-ph/9810374],
  M.~Bastero-Gil, G.L.~Kane and S.F.~King,
  Phys.\ Lett.\  B {\bf 474}, 103 (2000)
  [hep-ph/9910506].

\bibitem{Abe:2007kf}
  H.~Abe, T.~Kobayashi and Y.~Omura,
  Phys.\ Rev.\  D {\bf 76}, 015002 (2007)
  [hep-ph/0703044].

\bibitem{compressed}
S.P.~Martin,
  Phys.\ Rev.\  D {\bf 75}, 115005 (2007)
  [hep-ph/0703097].

\bibitem{Ellis:1984bm}
  J.R.~Ellis, C.~Kounnas and D.V.~Nanopoulos,
  Nucl.\ Phys.\  B {\bf 247}, 373 (1984).

\bibitem{Ellis:1985jn}
  J.R.~Ellis, K.~Enqvist, D.V.~Nanopoulos and K.~Tamvakis,
  Phys.\ Lett.\  B {\bf 155}, 381 (1985).

\bibitem{Drees:1985bx}
  M.~Drees,
  Phys.\ Lett.\  B {\bf 158}, 409 (1985).

\bibitem{Anderson:1996bg}
  G.~Anderson, C.H.~Chen, J.F.~Gunion, J.D.~Lykken, T.~Moroi and Y.~Yamada,
  ``Motivations for and implications of non-universal GUT-scale boundary
  conditions for soft SUSY-breaking parameters,''
{\it In the Proceedings of 1996 DPF / DPB Summer Study on New Directions 
for High-Energy Physics (Snowmass 96)}, 
  [hep-ph/9609457].

\bibitem{SUSYGUTS}
S.~Dimopoulos, S.~Raby and F.~Wilczek,
  Phys.\ Rev.\  D {\bf 24}, 1681 (1981).
S.~Dimopoulos and H.~Georgi,
  Nucl.\ Phys.\  B {\bf 193}, 150 (1981).
L.E.~Ibanez and G.G.~Ross,
  Phys.\ Lett.\  B {\bf 105}, 439 (1981).
N.~Sakai,
  Z.\ Phys.\  C {\bf 11}, 153 (1981).
A.H.~Chamseddine, R.L.~Arnowitt and P.~Nath,
  Phys.\ Rev.\ Lett.\  {\bf 49}, 970 (1982).
M.B.~Einhorn and D.R.T.~Jones,
  Nucl.\ Phys.\  B {\bf 196}, 475 (1982).
W.J.~Marciano and G.~Senjanovic,
  Phys.\ Rev.\  D {\bf 25}, 3092 (1982).
R.L.~Arnowitt, A.H.~Chamseddine and P.~Nath,
  Phys.\ Rev.\ Lett.\  {\bf 50}, 232 (1983).

\bibitem{SU5GUT}
  H.~Georgi and S.~L.~Glashow,
  Phys.\ Rev.\ Lett.\  {\bf 32}, 438 (1974).

\bibitem{Slansky}
  R.~Slansky,
  Phys.\ Rept.\  {\bf 79}, 1 (1981).

\bibitem{Anderson:1999uia}
  G.~Anderson, H.~Baer, C.h.~Chen and X.~Tata,
  Phys.\ Rev.\  D {\bf 61}, 095005 (2000)
  [hep-ph/9903370].

\bibitem{Huitu:1999vx}
  K.~Huitu, Y.~Kawamura, T.~Kobayashi and K.~Puolamaki,
  Phys.\ Rev.\  D {\bf 61} (2000) 035001
  [hep-ph/9903528].

\bibitem{Gunion:1999jr}
  J.F.~Gunion and S.~Mrenna,
  Phys.\ Rev.\  D {\bf 62}, 015002 (2000)
  [hep-ph/9906270].

\bibitem{Corsetti:2000yq}
  A.~Corsetti and P.~Nath,
  Phys.\ Rev.\  D {\bf 64}, 125010 (2001)
  [hep-ph/0003186].

\bibitem{Chattopadhyay:2001va}
  U.~Chattopadhyay, A.~Corsetti and P.~Nath,
  Phys.\ Rev.\  D {\bf 66}, 035003 (2002)
  [hep-ph/0201001].

\bibitem{Pallis:2003aw}
  C.~Pallis,
  Nucl.\ Phys.\  B {\bf 678}, 398 (2004)
  [hep-ph/0304047].

\bibitem{Chattopadhyay:2003yk}
  U.~Chattopadhyay and D.P.~Roy,
  Phys.\ Rev.\  D {\bf 68}, 033010 (2003)
  [hep-ph/0304108].

\bibitem{Profumo:2003em}
  S.~Profumo and C.E.~Yaguna,
  Nucl.\ Phys.\  B {\bf 681}, 247 (2004)
  [hep-ph/0307225].

\bibitem{Huitu:2005wh}
  K.~Huitu, J.~Laamanen, P.N.~Pandita and S.~Roy,
  Phys.\ Rev.\  D {\bf 72}, 055013 (2005)
  [hep-ph/0502100].

\bibitem{Baer:2005zc}
  H.~Baer, A.~Mustafayev, E.~K.~Park and S.~Profumo,
  JHEP {\bf 0507}, 046 (2005)
  [hep-ph/0505227].

\bibitem{Baer:2005jq}
  H.~Baer, T.~Krupovnickas, A.~Mustafayev, E.~K.~Park, S.~Profumo and X.~Tata,
  JHEP {\bf 0512}, 011 (2005)
  [hep-ph/0511034].

\bibitem{King:2007vh}
  S.F.~King, J.P.~Roberts and D.P.~Roy,
  JHEP {\bf 0710}, 106 (2007)
  [hep-ph/0705.4219].

\bibitem{Ananthanarayan:2007fj}
  B.~Ananthanarayan and P.N.~Pandita,
  Int.\ J.\ Mod.\ Phys.\  A {\bf 22}, 3229 (2007)
  [hep-ph/0706.2560].

\bibitem{compressed2}
S.P.~Martin,
  Phys.\ Rev.\  D {\bf 76}, 095005 (2007)
  [hep-ph/0707.2812],
  Phys.\ Rev.\  D {\bf 78}, 055019 (2008)
  [hep-ph/0807.2820].

\bibitem{Baer:2007uz}
  H.~Baer, A.~Box, E.K.~Park and X.~Tata,
  JHEP {\bf 0708}, 060 (2007)
  [hep-ph/0707.0618].

\bibitem{Bhattacharya:2007dr}
  S.~Bhattacharya, A.~Datta and B.~Mukhopadhyaya,
  JHEP {\bf 0710}, 080 (2007)
  [hep-ph/0708.2427].

\bibitem{Baer:2007xd}
  H.~Baer, A.~Mustafayev, H.~Summy and X.~Tata,
  JHEP {\bf 0710}, 088 (2007)
  [hep-ph/0708.4003].

\bibitem{Huitu:2008sa}
  K.~Huitu, R.~Kinnunen, J.~Laamanen, S.~Lehti, S.~Roy and T.~Salminen,
  Eur.\ Phys.\ J.\  C {\bf 58}, 591 (2008)
  [hep-ph/0808.3094].

\bibitem{Bhattacharya:2008qu}
  S.~Bhattacharya, A.~Datta and B.~Mukhopadhyaya,
  Phys.\ Rev.\  D {\bf 78}, 115018 (2008)
  [hep-ph/0809.2012].

\bibitem{Chattopadhyay:2009fr}
  U.~Chattopadhyay, D.~Das and D.~P.~Roy,
  [hep-ph/0902.4568].

\bibitem{SO10GUT}
  H.~Georgi, Particles and Fields, {\it Proceedings of the APS
  Division of Particles and Fields}, ed.~C.~Carlson, p.~575 (1975). 
  H.~Fritzsch and P.~Minkowski,
  Annals Phys.\  {\bf 93}, 193 (1975).

\bibitem{flippedSU5}
  A.~De Rujula, H.~Georgi and S.L.~Glashow,
  Phys.\ Rev.\ Lett.\  {\bf 45}, 413 (1980).
%
  S.M.~Barr,
  Phys.\ Lett.\  B {\bf 112}, 219 (1982).
%
  I.~Antoniadis, J.R.~Ellis, J.S.~Hagelin and D.V.~Nanopoulos,
  Phys.\ Lett.\  B {\bf 194}, 231 (1987).

\bibitem{PatiSalam}
  J.C.~Pati and A.~Salam,
  Phys.\ Rev.\  D {\bf 10}, 275 (1974)
  [Erratum-ibid.\  D {\bf 11}, 703 (1975)].

\bibitem{Chakrabortty:2008zk}
  J.~Chakrabortty and A.~Raychaudhuri,
  Phys.\ Lett.\  B {\bf 673}, 57 (2009)
  [hep-ph/0812.2783].

\bibitem{Chamoun:2001in}
  N.~Chamoun, C.S.~Huang, C.~Liu and X.H.~Wu,
  Nucl.\ Phys.\  B {\bf 624}, 81 (2002)
  [hep-ph/0110332].

\bibitem{E6GUT}
  F.~G\"ursey, P.~Ramond and P.~Sikivie,
  Phys.\ Lett.\  B {\bf 60}, 177 (1976).

\bibitem{flippedSO10}
  T.W.~Kephart and N.~Nakagawa,
  Phys.\ Rev.\  D {\bf 30}, 1978 (1984).

\bibitem{trinification}
  Y.~Achiman and B.~Stech, p. 303, ``New Phenomena in Lepton-Hadron
  Physics", ed. D.E.C.~Fries and J.~Wess, Plenum, NY, (1979).
  A.~de Rujula et al., p.~88, 5th Workshop on Grand Unification,
  ed.~K.~Kang, et al., World Scientific, Singapore (1984).

\bibitem{SU5thresholds}
  J.~Hisano, H.~Murayama and T.~Goto,
  Phys.\ Rev.\  D {\bf 49}, 1446 (1994).
  K.~Tobe and J.D.~Wells,
  Phys.\ Lett.\  B {\bf 588}, 99 (2004)
  [hep-ph/0312159].

\end{thebibliography}
\end{document}